\documentclass{iopart}

\usepackage{iopams}
\usepackage{graphicx}

\newcommand{\aver}[1]{\left\langle #1 \right\rangle}
\def\bra#1{\left\langle #1 \right|}
\def\ket#1{\left| #1 \right\rangle}

\newcommand{\quantop}[1]{\mathcal{#1}}
\newcommand{\HH}{\quantop{H}}

\begin{document}

\title[Complex dynamics of photon entanglement]{Complex dynamics
of photon entanglement in two-mode Jaynes-Cummings model}

\author{M. Erementchouk, %
M. N. Leuenberger}

\address{NanoScience Technology Center and Department of Physics, University of Central
Florida, Orlando, FL 32826}
\ead{merement@mail.ucf.edu}

\begin{abstract}
We study the dynamics of the photon entanglement, $E_{\mathrm{N}}(t)$, for
the two-mode Jaynes-Cummings model in the few-photon case. The atomic
transitions associated with the photons with different polarizations are
assumed to be independent and, hence, the evolution of the ``$+$"- and
``$-$"-polarized photons is formally separable. However, due to the photons
indistinguishability such interaction still leads to entanglement of
initially disentangled states owing to the non-linear dependence of the
characteristic frequencies on the photon population numbers. The time
dependence of entanglement is the result of superimposing oscillations with
incommensurate frequencies. Therefore, $E_{\mathrm{N}}(t)$ is a
quasi-periodic function of time with the complex profile strongly depending
on the number of photons.
\end{abstract}


\maketitle

\section{Introduction}

The property of many-body quantum systems to be in entangled, or
inseparable, states is the object of significant constant interest
\cite{HORODECKI:2009:ID3826}. Entanglement is the essentially quantum
feature signifying the formation of new entities, the complexes of
particles. If one would try to separate an individual particle from
such complex by performing a measurement one would inevitably affect
the states of other particles in the complex even if the direct
interaction is absent.

Because of their highly unusual, from the classical point of view,
properties the entangled states attract attention from both purely
scientific and application points of view. As a result, as a problem
of special importance there is the problem of producing entangled
states. For example, nowadays the most developed and widely used
method of generating the entangled photons is the parametric down
conversion \cite{Shih,Ou}, which is based on the two-photon radiative
decay of excited states. This method, however, suffers from intrinsic
limitations --- very low yield and rescaling the wavelength of the
emitted photons \cite{Bouwmeester,MandelWolf,Tanzilli}. Therefore,
there is the constant search of alternative sources of entangled
light, which motivates thorough investigating of the physics of
entangling.

The analysis of entangling photons requires different approach
comparing to that of canonical quantum mechanical systems, say,
qubits. First, one has to take into account indistinguishability of
the photons, which imposes the severe restrictions on available set
of states. Second, the interaction of the quantized electromagnetic
field with medium excitations is accompanied by the processes of
absorption and re-emission, hence the description should naturally
incorporate the fact that the number of particles is not a constant.
These circumstances suggest that the most suitable framework for
description of the photon entanglement is provided by the quantum
field theory. Within this framework the entanglement is characterized
using the single-particle density matrix (SPDM)
\begin{equation}\label{eq:SDM_def}
  K_{kq}(t) = \aver{a^\dagger_k(t) a_q(t)},
\end{equation}
where $k$ and $q$ enumerate the modes of the field. For example, for
the situation considered in the present paper $k$ and $q$ run over
``$+$" and ``$-$" polarizations.

Throughout the paper we incorporate the time dependence into the
Heisenber representation of the field operators $a_k(t) = \exp(\rmi
\HH t) a_k \exp(-\rmi \HH t)$ with $\HH$ being the Hamiltonian of the
system, therefore, the average in formulas similar to
(\ref{eq:SDM_def}) is taken with respect to the initial state
$\aver{\ldots} = \bra{\psi(t=0)}\ldots \ket{\psi(t=0)}$. We assume
that initially the system is in a pure state and that the decohering
processes are absent, therefore entanglement manifests itself in
mixed single-particle states, that is the SPDM has rank (or the
Schmidt number) larger than one
\cite{PASKAUSKAS:2001:ID3811,WANG:2005:ID3868}. More specifically the
entanglement can be quantified by the von Neumann entropy of the SPDM
\begin{equation}\label{eq:vNentropy}
  E_{\mathrm{N}} = - \sum_i \widetilde{\rho}_i \log(\widetilde{\rho}_i),
\end{equation}
where $\widetilde{\rho}_i$ are the normalized eigenvalues of the
SPDM, so that $\sum_i \widetilde{\rho}_i \equiv 1$.

In the present paper we study the time dependence of entanglement of
initially disentangled few-photon states in the two-mode
Jaynes-Cummings model. This model closely corresponds to the dynamics
of entanglement of the photons in the initially pumped cavity with a
single atom admitting transitions with different helicities. The
absorption and re-emission of photons with different polarizations is
assumed to be completely independent. From the canonical point of
view this may seem to be similar to a many-body system without
interaction, where entanglement is an integral of motion, that is
initially disentangled states remain disentangled. However, as we
will show, due to the complex character of the photons states the
interaction with the atom leads to entangling states with different
polarizations with non-trivial time dependence of
$E_{\mathrm{N}}(t)$.

\section{Photon entanglement in the context of Schwinger's model}

The most convenient framework for the description of the photon
entanglement is provided by Schwinger's model of angular momentum
\cite{SAKURAI:1994:ID3942}. The components of the operator of angular
momentum are defined in terms of the photons creation and
annihilation operators as
\begin{equation}\label{eq:ang_momentum_ops}
\eqalign{
  \quantop{J}_x = \frac 1 2 (a_+^\dagger a_- + a_-^\dagger a_+), \cr
  \quantop{J}_y = \frac 1 {2i} (a_+^\dagger a_- - a_-^\dagger a_+), \cr
  \quantop{J}_z = \frac 1 2 (a_+^\dagger a_+ - a_-^\dagger a_-).
}
\end{equation}
Using these operators one can characterize the photons states by the
total angular momentum $j$ and its projection $m$ instead of the
population numbers, $n_+$ and $n_-$, of the states with ``$+$"- and
``$-$"-polarizations. So, one has $\ket{n_+,n_-}
=\ket{j,m}_{\mathrm{S}}$ with the relations $j = (n_+ + n_-)/2$ and
$m=(n_+ - n_-)/2$.

Expressing the SPDM in terms of the mean values of the operator of
the angular momentum one finds that the normalized eigenvalues of the
SPDM can be expressed as
\begin{equation}\label{eq:eigen_SPDM_angular}
\widetilde{\rho}_{1,2} = \frac 1 2 (1- \widetilde{J}),
\end{equation}
where $\widetilde{J} = |\aver{\mathbf{J}}|/ j = 2
|\aver{\mathbf{J}}|/N$ with $N = \Tr[K]$ being the total number of
the photons. Thus, the entanglement can be written as $E_{\mathrm{N}}
= F(\widetilde{J})$, where
\begin{equation}\label{eq:def_F_ent}
  F(x) = \frac 1 2 \sum_{n=1,2} \left[1 + (-1)^n x \right] \log_2\left[1 + (-1)^n x \right].
\end{equation}
In particular, disentangled states are the states with the maximum
magnitude of the average angular momentum $|\aver{\mathbf{J}}| = j$
and completely entangled ones are those with $|\aver{\mathbf{J}}| =
0$. It is interesting to compare this representation with the
description of entanglement in the canonical quantum mechanical
setup. Considering the two-photon case, $N=2$, one can find that the
total value of the angular momentum is directly related to the
concurrence \cite{WOOTTERS:1998:ID3357} $|C|^2 = 1 -
\widetilde{J}^2$.

Such description of entanglement allows a simple parametrization of
disentangled states. This is especially convenient for the problem
with initially disentangled photons, which naturally appears when one
considers the free dynamics of initially pumped cavity with an atom.
The parametrization is based on the facts that for the disentangled
states and the disentangled states only one has $|\aver{\mathbf{J}}|
= j$ and that the average angular momentum $\aver{\mathbf{J}}$
transforms under rotations as a three-dimensional vector. Hence, any
disentangled state can be turned by the rotations into
$\ket{j,j}_{\mathrm{S}}$, or, in other words, any disentangled state
can be presented as
\begin{equation}\label{eq:disent_inter}
  \ket{\psi(\beta_1,\beta_2,\beta_3)} = \exp(-\rmi \quantop{J}_z \beta_1)
  \exp(-\rmi \quantop{J}_y \beta_2)\exp(-\rmi \quantop{J}_z \beta_3) \ket{j,j}_{\mathrm{S}},
\end{equation}
where $\beta_{1,2,3}$ are the Euler angles. For the case of our main
interest in the present paper only rotations by angle $\beta_2$
produce the states with different dynamics of entanglement,
therefore, in the following consideration we limit our attention to
$\ket{\psi(\beta)} = \exp(-\rmi \quantop{J}_y \beta)
\ket{j,j}_{\mathrm{S}}$.

\section{Time dependence of entanglement}

The dynamics of the two-mode Jaynes-Cummings model
\cite{SHORE:1993:ID3860} is governed by the Hamiltonian, which we
write down in terms of the creation and annihilation operators
\begin{equation}\label{eq:jc_two-mode}
\HH = \sum_k \epsilon^{(\mathrm{p})}_k a_k^\dagger a_k +
    \sum_\kappa \epsilon^{(\mathrm{e})}_\kappa c_\kappa^\dagger c_\kappa
   + \HH_+ + \HH_-.
\end{equation}
Here the first two terms describe the dynamics of the free field and
the free atom, respectively. Here $k$ runs over the photon
polarizations, $+$ and $-$, and $\kappa$ takes the values from the
set of the electron states $\{\mathrm{g}\uparrow,
\mathrm{g}\downarrow, \mathrm{e}\uparrow, \mathrm{e}\downarrow\}$
where $\mathrm{g}$ and $\mathrm{e}$ stand for the ground and excited
atom levels and $\uparrow$, $\downarrow$ are the electron spins.

The interaction between the atom and photons preserves the helicity,
that is the excitation of the electron state with the spin down at
the ground level, $\ket{\mathrm{g}\downarrow}$, into the spin up
state at the excited level $\ket{\mathrm{e}\uparrow}$ occurs through
the absorption of ``$+$"-polarized photon and so on. The interaction
is described by the Hamiltonians $\HH_k = \omega_k a_k
\sigma^\dagger_k + \mathrm{h.c.}$, where $\sigma^\dagger_+ =
c_{\mathrm{e}\uparrow}^\dagger c_{\mathrm{g}\downarrow}$,
$\sigma^\dagger_- = c_{\mathrm{e}\downarrow}^\dagger
c_{\mathrm{g}\uparrow}$ and $\omega_\pm$ are the respective Rabi
frequencies.

The interplay between different characteristic frequencies
determining the dynamics of the system leads to the complex time
dependence of the amplitudes. In order to concentrate on the main
features, we adopt the resonant approximation and set
$\epsilon^{(\mathrm{e})}_\kappa = \epsilon^{(\mathrm{p})}_k = 0$.
Additionally we assume that the symmetry between transitions with
different helicities is not broken so that $\omega_+ = \omega_- =
\omega_{\mathrm{R}}$.

In order to describe the time dependence of entanglement we use the
explicit form of the Heisenberg representation for the photon
operators $a_k(t) = \exp(\rmi t \quantop{H}_k) a_k \exp(-\rmi t
\quantop{H}_k)$. Taking into account the separability of the dynamics
we have \cite{Scully_QO}
\begin{equation}\label{eq:ev_op}
  a_k(t) = e^{i \quantop{C}_k t} \left\{
    \left[\cos\left(\bar{\quantop{C}}_k t\right) - \rmi \quantop{C}_k
    \frac{\sin\left(\bar{\quantop{C}}_k t\right)}{\bar{\quantop{C}}_k}
    \right] a_k - \rmi \frac {\sin\left(\bar{\quantop{C}}_k t\right)}{\bar{\quantop{C}}_k} \sigma_k
  \right\},
\end{equation}
where $\quantop{C}_k^2 = \omega_{\mathrm{R}} a_k^\dagger a_k$ and
$\bar{\quantop{C}}_k^2 = \omega_{\mathrm{R}} (a_k^\dagger a_k + 1)$.
Taking into account that initially the atom is not excited the
initial state of the system is presented as
\begin{equation}\label{eq:jc_initial_state}
  \ket{\Psi(0)} = \ket{\psi(\beta)} \ket{0}_e,
\end{equation}
where $\ket{\psi(\beta)}$ is the disentangled photon state obtained
by rotations from the state $\ket{j,j}_{\mathrm{S}}$ [see
(\ref{eq:disent_inter})] and $\ket{0}_e$ denotes the state of the
atom with both electrons at the ground level. It should be noted with
this regard that the second term in (\ref{eq:ev_op}) does not
contribute to $K_{kq}(t)$ because it vanishes while acting on the
initial state of the atom.

First, we consider $\aver{\quantop{J}_z(t)}$. Taking into account the
identity $f\left(\bar{\quantop{C}}_k\right) a_ k = a_k
f\left({\quantop{C}}_k\right)$ we find
\begin{equation}\label{eq:jc_Jz}
 \aver{\quantop{J}_z(t)} = \aver{\quantop{J}_z} +
    \frac 1 4 \aver{\left[\cos(2 \quantop{C}_+ t) - \cos(2 \quantop{C}_- t)\right]}.
\end{equation}
Here and below the average is taken over the initial photon state
$\aver{\ldots} = \bra{\psi(\beta)} \ldots \ket{\psi(\beta)}$. The
second term in (\ref{eq:jc_Jz}) has the maximum magnitude in
completely polarized case, that is when $\aver{\quantop{J}_z}=j$,
when it has the form of oscillations with the frequency $2 N
\omega_{\mathrm{R}}$ due to absorption and re-emission of the single
photon by the atom signifying the entanglement of photons and atomic
states. The magnitude of this term monotonously decreases with
$\beta$ and vanishes identically in the unpolarized case, when
$\aver{\quantop{J}_z}=0$.

It should be noted that even in the polarized case the relative
contribution of the last term in (\ref{eq:jc_Jz}) is relatively small
$\propto 1/N$ and thus one can consider $\aver{\quantop{J}_z}$ as
imposing the limitation from above on the maximum value of achievable
entanglement. Therefore we concentrate on the unpolarized case when
$\aver{\quantop{J}_z(t)} \equiv 0$ and the time dependence of
entanglement is only due to the transversal component of the angular
momentum $|\aver{\mathbf{J}(t)}|^2 = \aver{\quantop{J}_x(t)}^2 +
\aver{\quantop{J}_y(t)}^2$.

\begin{figure}
  \includegraphics[width=5in]{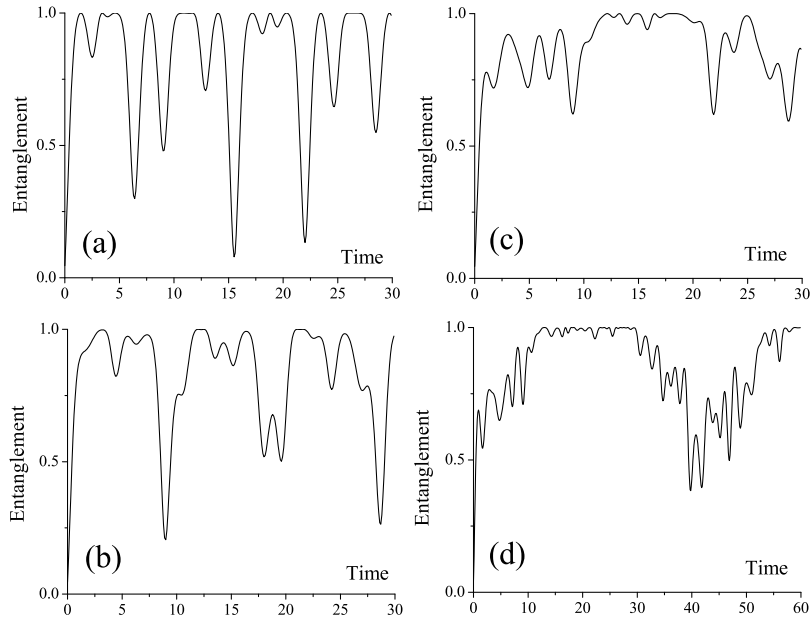}
  \caption{Time dependence of entanglement of the initially
  disentangled $N$ photons: (a) $N=2$, (b) $N=3$, (c) $N=4$ and (d) $N=5$
  (notice the different total duration).
  The time is measured in $1/\omega_{\mathrm{R}}$. }\label{fig:ent}
\end{figure}

We present in figure~\ref{fig:ent} the results of numerical
evaluation of entanglement time dependence calculated according to
equation (\ref{eq:def_F_ent}). First of all we would like to
emphasize that the entanglement reaches the maximum value even in the
case when $N > 2$, when the sole entanglement with the atomic state
is not sufficient to support $E_{\mathrm{N}} = 1$. In order to
provide a qualitative explanation of the origin of entanglement let
us consider a two-photon state. In
the basis of population numbers any such state has the form
\begin{equation}\label{eq:two_boson_pop}
  |\psi \rangle = \alpha_{2,0} \ket{2,0} + \alpha_{1,1} \ket{1,1}
  + \alpha_{0,2} \ket{0,2},
\end{equation}
where $\alpha_{n_+, n_-}$ are the respective time-dependent
amplitudes. Using this representation we find for the concurrence
\begin{equation}\label{eq:conc_alphas}
  C = 2 \alpha_{2,0}\alpha_{0,2} - \alpha_{1,1}^2.
\end{equation}
The initial condition is determined by the requirement for the state
to be disentangled, thus $C(t=0)= 0$. For a free field the
characteristic frequencies of the amplitudes $\alpha_{n_+, n_-}$ are
\emph{linearly} proportional to the total number of photons and, as a
result, the time dependence factors out in (\ref{eq:conc_alphas})
yielding $C(t)=0$, that is the state remains disentangled. However,
as can be seen from Heisenberg representation (\ref{eq:ev_op}), the
interaction with the atom essentially modifies the dependence of the
characteristic frequencies on the number of photons, which becomes
$\propto \sqrt{N}$. Thus the time dependence of the first and the
second terms in (\ref{eq:conc_alphas}) is determined by the
incommensurate frequencies $\Omega_1 / \Omega_2 \sim \sqrt{2}$, which
inevitably leads to the phase desynchronizing and to the violation of
the condition $C= 0$.

The second important consequence of the involvement of incommensurate
frequencies into the dynamics is the complex (quasi-periodic) profile
of $E_{\mathrm{N}}(t)$, which dominates the time dependence of
entanglement after the initial regular raise with the typical time
scale $\sim 1/\omega_{\mathrm{R}}$. This makes $E_{\mathrm{N}}(t)$ to
some extent unpredictable and imposes strong requirements with regard
to the control of the life time of the photons inside the cavity and
to the initial conditions. The addition of a single photon
drastically changes the value of entanglement at any particular
instant beyond the period of initial raise. It should be noted,
however, that when the number of photons in the cavity increases the
contribution of particular frequencies becomes less important and the
profile of $E_{\mathrm{N}}(t)$ changes toward some general regular
shape. The initial stage of the transformation can be seen by
comparing figures~\ref{fig:ent}c and \ref{fig:ent}d. The detailed
analysis of the regular profile in the limit $N \gg 1$ will be
provided elsewhere. Here we would like to note that the normalized
value of the average angular momentum can be presented in this limit
as a periodic sequence of Gaussian bumps with the characteristic
period $\sim N^{3/2}/ \omega_{\mathrm{R}}$ and the typical
entanglement time $\sim N/\omega_{\mathrm{R}}$.

\section{Conclusion}

We have considered the time dependence of entanglement of initially
disentangled few-photons states within the two-mode Jaynes-Cummings
model. The processes of absorption and re-emission of photons with
different polarizations are assumed independent and, therefore, the
dynamics of ``$+$"- and ``$-$"-polarized photons is completely
separable. Applying the standard description of entanglement
straightforwardly this may seem to imply that the states with
different polarizations will stay disentangled. However, the
indistinguishability of the photons leads to the necessity to apply
the standard picture with the special care. Indeed, except in the
simplest case, when all photons have the same polarization, each
photon is always in the superposition of the states and, therefore,
is always affected by ``both parts" of the dynamics. As a result, in
order to make a conclusion regarding entanglement one has to consider
closely the evolution of the relation between the amplitudes.

We have studied the photon entanglement using the formalism of the
single-particle density matrix. We have established the relation
between the entanglement and the magnitude of the average angular
momentum defined with the help of Schwinger's model. Using this
relation and the exact Heisenberg representation of the photon
creation and annihilation operators we have calculated the time
dependence of entanglement $E_{\mathrm{N}}(t)$. We have shown that
$E_{\mathrm{N}}(t)$ has two important features. The interaction with
the single atom may lead to complete entanglement $E_{\mathrm{N}} =
1$. This is related to the fact that the characteristic frequencies
determining the time evolution of the photon amplitudes depend
\emph{non-linearly} on the population numbers. In turn, the
superposition of oscillations with incommensurate frequencies results
in quasi-periodic $E_{\mathrm{N}}(t)$ with complex profile,
especially in the few-photon case $N \sim 1$. The specific form of
the profile drastically depends on the number of particles and
transforms into a regular pattern in the limit $N \gg 1$,

\section*{References}


\end{document}